\documentclass{article}

\usepackage[nonatbib, preprint]{neurips_2021}

\usepackage[utf8]{inputenc} 
\usepackage[T1]{fontenc}    
\usepackage{hyperref}       
\usepackage{url}            
\usepackage{booktabs}       
\usepackage{amsfonts}       
\usepackage{nicefrac}       
\usepackage{microtype}      
\usepackage{xcolor}         
\usepackage{microtype}
\usepackage{graphicx}
\usepackage{subfigure}
\usepackage{booktabs} 
\usepackage{tabularx, longtable}
\usepackage{hyperref}

\title{On Two XAI Cultures:\\ A Case Study of Non-technical Explanations in Deployed AI System}

\author{%
	Helen Jiang\thanks{Corresponding author, alternative and preferred email: helen.h.jiang@gmail.com} \\
	Department of Computer Science\\
	Georgia Institute of Technology \\
	Atlanta, GA  30332\\
	\texttt{hjiang328@gatech.edu} \\
	\And
	Erwen Senge \\
	Independent \\
	Montréal, QC H3A 0G4  \\
	\texttt{erwen@protonmail.com} \\

}

\begin{document}
	
	\maketitle
	
	\begin{abstract}
		Explainable AI (XAI) research has been booming, but the question ``\emph{To whom} are we making AI explainable?'' is yet to gain sufficient attention. Not much of XAI is comprehensible to non-AI experts, who nonetheless, are the primary audience and major stakeholders of deployed AI systems in practice. The gap is glaring: what is considered ``explained'' to AI-experts versus non-experts are very different in practical scenarios. Hence, this gap produced two distinct cultures of expectations, goals, and forms of XAI in real-life AI deployments\cite{towhom:preece2018stakeholders}.

		We advocate that it is critical to develop XAI methods for non-technical audiences. We then present a real-life case study, where AI experts provided non-technical explanations of AI decisions to non-technical stakeholders, and completed a successful deployment in a highly regulated industry. We then synthesize lessons learned from the case, and share a list of suggestions for AI experts to consider when explaining AI decisions to non-technical stakeholders. 
	\end{abstract}
	
	\section{Introduction} \label{sec:intro}
	While \cite{kim:xai:doshi2017towards} provided a working definition for interpretability (``ability to explain or to present in understandable terms to a human''), It is often used interchangeably with ``explainability''. By looking at much of XAI literature, it seems the XAI community was too ready to assume that such a human is an AI expert\cite{towhom:user:kirsch2017explain,towhom:notlay:bhatt2020explainable}. However, in real-life AI deployments, many humans are not AI experts, and want to understand decisions made by AI. In light of this reality, it is easy to see that AI explanations which AI experts understands, may well be inexplicable and incomprehensible to human non-experts. The gap between these two audiences of XAI is a problem. Non-experts --- users, application domain experts, regulatory officers, policymakers, etc. --- are nonetheless stakeholders\cite{towhom:preece2018stakeholders} in AI systems used in real-life, and not providing sufficient explanability to such a broad range of audience, is decidedly a neglected spot of AI community's research efforts. 
	
	Here we aim to raise more awareness on XAI targeting non-technical audience. By presenting a case study, we sketch a viable roadmap to bridge the two distinct needs for XAI, and invite the broader AI community to seek more solutions to bring along non-technical stakeholders in real-life AI deployments. 
	
	We first provide working definitions for the following terms that we use. A \textbf{\emph{non-technical}} or \textbf{\emph{lay audience}} for XAI is a generally well-educated (e.g. a Bachelor's degree) group, and while they are not well-versed in the demanding statistics and mathematics needed by AI experts, they share a foundation of numerical literacy. Similarly, a \textbf{\emph{non-technical explanation}} would be an explanation that a non-technical audience could readily understand, \emph{without intensive educational efforts} from AI experts. Conversely, a \textbf{\emph{technical explanation}} is an explanation with the rigor and sophistication that a \emph{technical} AI expert demands. An example of non-technical audience, would be financial compliance officers with Accounting and Law degrees: to them, presenting feature importance lists for generalized linear models, or a single decision tree with features, labels, and probabilities on each split, are non-technical explanations readily understood by them. On the other hand, if they receive SHAP summary plots and dependency plots, they would need extensive educational efforts from AI experts to understand. 
	
	We are aware that as \cite{mittelstadt2019explaining, towhom:lipton2018mythos} point out, the XAI community seems to have divergent definitions for what an ``explanation'' is and what ``explainability'' means, but this is out of scope of our paper. Instead, here we focus on the \textbf{\emph{end goals}} of explanations, in the context of real-life AI deployments. 
	
	An explanation should \emph{enable} at least one of the three following items: 
	\begin{itemize}
		\item Make decisions;
		\item Take actions;
		\item increase understandings on how AI made its decisions. 
	\end{itemize}
	For example, compliance officers deciding if a lending model discriminates against protected groups, is an end goal that XAI explanations could empower.
	
	
	\section{Related Work} \label{section:relatedwork}
	
	With roots in expert systems dating back to the 1970s and continued into the 1990s\cite{ogxai:shortliffe1974mycin, ogxai:sotos1990mycin, ogxai:swartout1993explanation}, XAI research enjoyed a resurgence, with high-profile call-to-actions from experts\cite{darpa:xai2016} and the popular press\cite{xai:press} alike. However, the ``inmates are running the asylum'' problem is as visible in current XAI research as it has been in general software systems\cite{inmates:cooper2004inmates}: \cite{inmates:miller2017explainable,inmates:miller2019explanation,towhom:notlay:bhatt2020explainable} have already shown plenty of examples. Much of XAI research build explanations for those \textit{who are already AI experts}. Overall, little regard is given to non-AI-expert stakeholders who need AI explanations, and have different goals for those explanations. This produces a large gap between technical experts' version of XAI, and what non-technical audiences expect of XAI. Researchers have started to explore XAI working in tandem with social and behavioral sciences\cite{inmates:miller2019explanation} as well as philosophy\cite{mittelstadt2019explaining}, but XAI community still falls behind in explaining and communicating AI decisions to non-technical stakeholders who try to understand deployed AI systems in real-life. In \cite{towhom:framework:littleaudience:sokol2020explainability}'s excellent framework to assess XAI, ``target audience'' is a very small part of consideration, and \cite{towhom:user:kirsch2017explain, towhom:bohlender2019towards} focus on theories of explainable systems and are amiss in practical and solutions. Even XAI user studies\cite{towhom:mohseni2018human, towhom:narayanan2018humans, towhom:tomsett2018interpretable} did not specify with sufficient details, \emph{who} the users or study participants actually are, before showing AI explanations to them.  
	
	One example of this phenomenon, would be stakeholders in highly regulated industries such as banking, lending, insurance, and finance. They have very stronge incentives to understand AI decisions before deploying AI systems at scale, largely to comply with regulatory requirements\cite{fairlending:dontriskit, letter:warren:algo:fintech,aireg:fdic:americanbanker}. 
	\footnote{In the U.S., AI applications, their decisions, explanations, and documentations may be required, and/or subject to regulatory compliance oversight and reviews, by the following legislation and/or statutes: Civil Rights Acts of 1964 and 1991, Equal Credit Opportunity Act, Fair Housing Act, Fair Credit Reporting Act, Americans with Disabilities Act, Rehabilitation Act of 1973, Genetic Information Nondiscrimination Act, Health Insurance Portability and Accountability Act, Age Discrimination in Employment Act, Federal Reserve SR 11-7, and Article 22 of European Union General Data Protection Regulation, etc.} However, much of XAI work has been directed to technical audiences with deep knowledge in mathematics and computer science, instead of a lay audience who's generally well-educated but not necessarily technical\cite{compliance:jobreq:onet,compliance:jobreq1,compliance:bls,compliance:eeoc}. 
	
	\section{Case Study: XAI in Consumer Lending for Compliance \& Legal Professionals} \label{section:casestudy}
	This real-life case study is based on one author's 8-month internship at the consumer lending business unit of a financial services firm operating in the United States, until June 2021. The business unit has a dedicated Machine Learning (ML) team of more than 12 researchers, data scientists, and engineers, and the case study focuses on a binary classification ML model they deployed. The model makes underwriting decisions: consumers apply for lines of credit, and the model will decide whether the application is approved or denied. In this case, the ML team leveraged both technical and non-technical XAI approaches to communicate the model's ethical and legal implications to an internal, non-technical audience, largely composed of compliance and legal professionals. 
	
	\subsection{Target Audiences \& Their Expectations for AI Explanations}
	The ML team collaborated with the Compliance team and the Legal team, with their priorities and end goals for having AI explanations summarized in table \ref{team-table}. Both teams' members have at least Bachelor degrees in Business, Finance, or Liberal Arts disciplines, but none in science, technology, engineering, or mathematics, and none has worked in these domains. All but two have post-graduate law degrees. All of them could correctly (1) identify and give examples describing statistical outliers; (2) draw the shape of a ``bell curve''; and (3) conceptually understand what standard deviation measures. Yet they are not familiar with those terms' mathematical expressions and statistical caveats. Based on this information, we concluded that these teams all fit into our working definition of a \emph{non-technical} audience. 
	Focusing on the compliance and legal team, we discovered their needs for AI explanations:
	\begin{itemize}
		\item The explanations should be statistically rigorous as possible but still readily within their understanding;
		\item The explanations will be used to decide on a course of action (e.g. reverse a denial decision), or form an opinion (e.g. whether the model complies with regulatory standards);
		\item The explanations would need to be accompanied by documentations, precise quantitative outputs, and back tests.
	\end{itemize}

	\begin{table*}[t] 
		\caption{Overview of Stakeholder Teams in new ML Model's Deployment}
		\label{team-table}
		\vskip 0.15in
		\begin{center}
			\begin{tabular}{|p{1.8cm}|p{7cm}|p{7cm}|}
				\toprule
				TEAM  & TOP PRIORITY & GOAL FOR XAI  \\
				\midrule		
				ML & Model results in higher approval and lower default rates than existing underwriting rules & To debug, tune, and remedy model behaviors and features \\
				\midrule
				Compliance & Model complies with relevant consumer lending regulations, e.g. fair lending, equal credit & Evidence that model does not discriminate against protected groups \\
				\midrule
				Legal     & Model should cover all scenarios included or implied by law  & To evaluate if model is reasonably complete in legal aspects  \\
				
				\bottomrule
			\end{tabular}
			
		\end{center}
		\vskip -0.1in
	\end{table*}

	\subsection{Tools \& Procedures}
	For each development iteration, both Compliance and Legal received of documentations and model explanations. Among them, these documents were most frequently consulted: 
	
	\begin{enumerate}
		\item A summary of disparate impact (DI) for protected groups (See table \ref{sample-table} for an example); 
		\item A list of original features in the model; 
		\item \textbf{Simplified} forms of a number of technical XAI approaches, e.g. local interpretable model-agnostic explanations (LIME), individual conditional expectation (ICE) plots, partial dependence plots, decision tree surrogates, and Shapley explanations, etc.; 
		\item A list of the model's relative global feature importance;
		\item In addition, the Compliance and Legal teams would also receive a list of ``\textbf{corner cases}'' from back test. These ``corner cases'' are a mix of randomly selected individual applications, and applications with decisions overturned by a manual credit review team. 
	\end{enumerate}
	
	To produce the DI, the ML team adapted the well-known ratio representation of approval rates between different groups, to report and evaluate model fairness. Because lending functions are prohibited from collecting protected group information (e.g. race, gender), the ML team used external data sources to infer protected group status, which led to non-integer counts of approval and total in table \ref{sample-table}. 
	
	If Compliance and Legal concluded that some features constitute regulatory risk, or any protected group was disparately impacted, the ML team would remedy the model. To accomplish this, ML team used technical XAI methods to debug and remedy. For example, if a global tree surrogate of low error measure (e.g. RMSE, MAPE) showed an unusual split for a sub-population, the team might consider using tree SHAP (SHapley Additive explanations) to get locally accurate feature contributions, and to generate adverse action codes required by regulations. After the ``fixed'' model was tested to satisfaction by the ML team, a \emph{simplified} version of technical XAI approaches would be created for compliance and legal to review, and the evaluation and remedy cycle with Compliance and Legal would restart.

	\begin{table}[t]
		\caption{Example summary disparate impact, presented to Compliance and Legal teams. Numbers are fictional, and are rounded to 2 digits after decimal. IPG = Inferred Protected Group. }
		\label{sample-table}
		\vskip 0.15in
		\begin{center}
			\begin{small}
				\begin{sc}
					\begin{tabular}{|p{1.2cm}|p{1.5cm}|p{1.8cm}|p{1cm}|p{1cm}|} 
						\toprule
						Groups & Approval & Total & Rate & Impact Ratio \\
						\midrule
						Baseline & 89218.48 & 1,246,840.72 & 0.07 & 1.00 \\
						IPG 1 & 93071.22 & 1,495,032.24 & 0.06 & 0.87 \\
						IPG 2 & 23953.79 & 288,582.94 & 0.08 & 1.16 \\
						...  & ...       &  ...  & ...          & ...   \\
						
						\bottomrule
					\end{tabular}
				\end{sc}
			\end{small}
		\end{center}
		\vskip -0.1in
	\end{table}

	\section{Lessons Learned} \label{section:2cultures}
	We'd like to share some key practical lessons we learned from the case study. 
	\begin{itemize} \label{suggestionlist}
		
		\item \emph{Know the audience}: In more formal terms, this is called ``user modeling'', a long-established research technique in human-computer interactions (HCI)\cite{usermodel:rich1979user,usermodel:kobsa2001generic,usermodel:fischer2001user}. Describe in detail the desired attributes of targeted users, e.g. users' motivations, prior knowledge, behavioral patterns, etc. so researchers have sufficient contexts to devise methods and conduct studies. In our case study, understanding compliance and legal teams' priorities and interests in how they want to utilize XAI, was key for the successful deployment. %
		
		\item \emph{Help stakeholders build mental models of XAI}: ``Mental models'' are well-known in psychology and HCI: they are internal representations of a situation, that provide predictive and explanatory power to their beholders\cite{mentalmodel:johnson1983mental,mentalmodel:norman1983some}. Helping non-technical audiences build reasonable expectations of XAI, and how XAI can or cannot help with their end goals, would enhance audiences' understanding and usage of AI explanations. 
		
		\item \emph{Tailor the communication}: XAI results that AI experts can understand, may be incomprehensible to non-technical audiences. To present explanations in ways that non-technical audiences can readily process with their own prior knowledge and domain expertise, would be a large step towards bridging technical and non-technical XAI. 
		
		\item \emph{Incorporate insights from social sciences and domain expertise}: Economics and psychology have long established accessible forms to present research to a generally well-educated but not necessarily expert audience. For example, ``disparate impact'' for different groups, which the ML team presented to the compliance team, is well-researched in psychology and economics\cite{air:evenoff1996cra,air:frame2004credit,air:ladd1998evidence, air:hough2001determinants,air:morris2001sample,air:philyfed:credit_score,air:ployhart2008diversity,air:avery2012does,air:de2007combining}, and accepted as a measure for fairness by practicing professionals. They are indeed more accessible to non-technical audiences than fairness measures used by AI experts. Similarly, building domain knowledge into XAI, e.g. specifying XAI approaches in medical domains\cite{holzinger2017we,domainexpert:ahmad2018interpretable, holzinger2019causability}, also helps making XAI more accessible to an extremely well-educated yet largely non-AI-expert audience. 
		
	\end{itemize}

	\section{Reflections \& Conclusions \& Work in Progress} \label{sec:future}
	\subsection{Reflections} 
	It came to our attention that even some prominent AI research make erroneous assumptions about well-established domain-specific practices in real-life. We conjecture this \textbf{disconnect} between AI research and AI applications, is the root cause of the gap between the two XAI cultures. For example, what \cite{cluless:ustun2019actionable} claimed to be ``widespread practices of machine learning,'' according to members of the legal and compliance team, is either non-compliant with industry regulations (e.g. using \textit{married} or \textit{has\_phd} as features), or likely outright illegal (e.g. using \textit{age}). Meanwhile, ``effects test''\cite{biznec:cfpb} and ``business necessity test''\cite{biznec:scotus:griggs} widely known by regulatory and compliance stakeholders, did not seem to have come to the attention of many well-intended AI researchers. 
	
	\subsection{Conclusion}
	We encourage AI researchers to engage and collaborate with, and pay more attention to non-technical stakeholders of deployed AI systems. We propose that we revisit the original DARPA XAI call: ``...human-in-the-loop techniques that developers can use..\emph{for more intensive human evaluations}''(emphasis added), while the ``human'' doing those ``intensive'' evaluations need not be AI experts.

	\subsection{Work in Progress} 
	We are aware that bridging the approaches, expectations, and mindsets between technical and non-technical AI explanations, is a tough task, and needs more research and discussions, not only in the XAI community, but also interdisciplinary collaborations from HCI\cite{hci:tullio2007works,hci:kulesza2013too, hci:design:zhu2018explainable,hci:guzdial2019friend,hci:liapis2019fusing}, social sciences, and behavioral sciences\cite{apa:kahneman1982judgment,apa:hilton1986knowledge,ss:rehder2003causal,ss:lombrozo2009explanation}. Currently, based on our case study and literature research, we are proposing a detailed workflow that put both technical and non-technical AI stakeholders at the center, for high-stake decisions in AI's deployment in real life.

	\bibliographystyle{acm}
	\bibliography{ai_culture}

\end{document}